\documentclass[10pt,conference]{IEEEtran}
\usepackage[cmex10]{amsmath}
\usepackage[short]{optidef}
\usepackage{color}
\usepackage{commath}
\usepackage{algorithm}
\usepackage{algorithmic}
\usepackage{textcomp}
\usepackage{gensymb}
\usepackage{multirow}
\usepackage{multicol}
\usepackage{mathtools}
\DeclareMathOperator*{\argmax}{arg\,max}

\ifCLASSINFOpdf
   \usepackage[pdftex]{graphicx}
\graphicspath{ {figures/} }
\else
\fi
\usepackage{graphicx}
\usepackage[font=footnotesize,caption=false,subrefformat=parens,labelformat=parens]{subfig}

\usepackage{cite}
\usepackage{amssymb}

\usepackage{amsthm}
\usepackage{pifont}

\usepackage{siunitx}


\newcommand\blfootnote[1]{%
		\begingroup
		\renewcommand\thefootnote{}\footnote{#1}%
		\addtocounter{footnote}{-1}%
		\endgroup
}

\title{Mobility Management for Cellular-Connected UAVs: A Learning-Based Approach}
\author{
   \IEEEauthorblockN{Md Moin Uddin Chowdhury\IEEEauthorrefmark{1}, Walid Saad\IEEEauthorrefmark{2}, and Ismail G{\"{u}}ven{\c{c}}\IEEEauthorrefmark{1}}
    \IEEEauthorblockA{\IEEEauthorrefmark{1}Department of Electrical and Computer Engineering, North Carolina State University, Raleigh, NC, USA}
       
    \IEEEauthorblockA{\IEEEauthorrefmark{2}Wireless@VT, Electrical and Computer Engineering Department, Virginia Tech, VA, USA\\
       Email: \{mchowdh,iguvenc\}@ncsu.edu,~walids@vt.edu}
}
\begin{document}
\pdfoutput=1
\maketitle
\begin{abstract}
The pervasiveness of the wireless cellular network can be a key enabler for the deployment of autonomous unmanned aerial vehicles (UAVs) in beyond visual line of sight scenarios without human control. However, traditional cellular networks are optimized for ground user equipment (GUE) such as smartphones which makes providing connectivity to flying UAVs very challenging. Moreover, ensuring better connectivity to a moving cellular-connected UAV is notoriously difficult due to the complex air-to-ground path loss model. In this paper, a novel mechanism is proposed to ensure robust wireless connectivity and mobility support for cellular-connected UAVs by tuning the downtilt (DT) angles of all the ground base stations (GBSs). By leveraging tools from reinforcement learning (RL), DT angles are dynamically adjusted by using a model-free RL algorithm. The goal is to provide efficient mobility support in the sky by maximizing the received signal quality at the UAV while also maintaining good throughput performance of the ground users. Simulation results show that the proposed RL-based mobility management (MM) technique can reduce the number of handovers while maintaining the performance goals, compared to the baseline MM scheme in which the network always keeps the DT angle fixed.
\end{abstract}
\section{Introduction}
The use of unmanned aerial vehicles (UAVs) in both civilian and commercial applications is a promising technology for next-generation cellular networks. Thanks to their ease of deployment, high flexibility, and the ability to increase network capacity, UAVs can be deployed in surveillance, remote sensing, package delivery, and disaster-relief applications. Hence, significant efforts in both academia and industry have been devoted to different aspects of UAV applications.
\blfootnote{This work is supported by NSF under grants CNS-1453678, CNS-1910153,  and CNS-1909372.}However, most of the above-mentioned UAV cases are operated by trained pilots.

\begin{figure}[t]
\centering
		\subfloat[Scenario: 1]{
			\includegraphics[width=.65\linewidth]{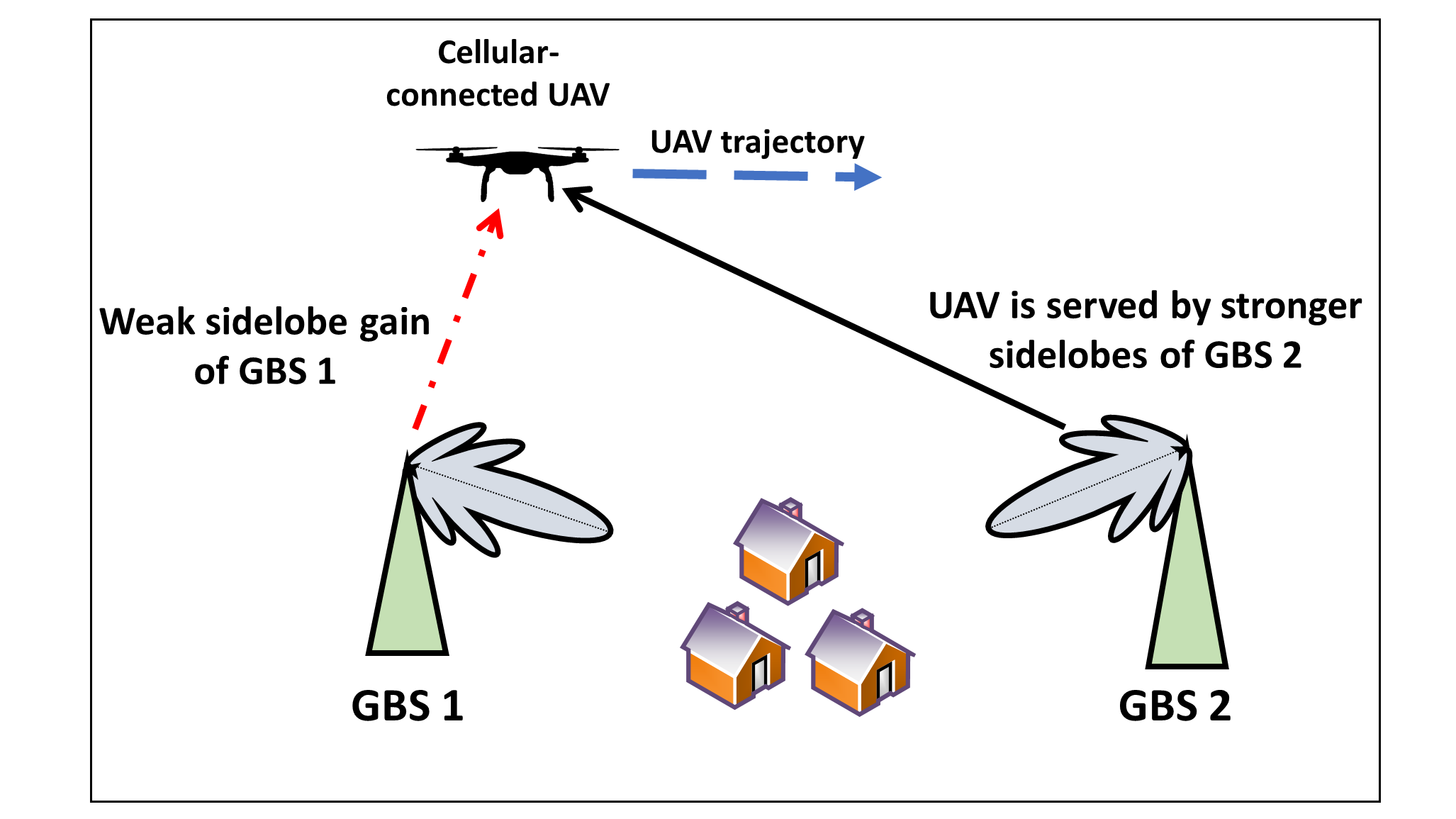}} \hfill
		\subfloat[Scenario: 2]{
			\includegraphics[width=.65\linewidth]{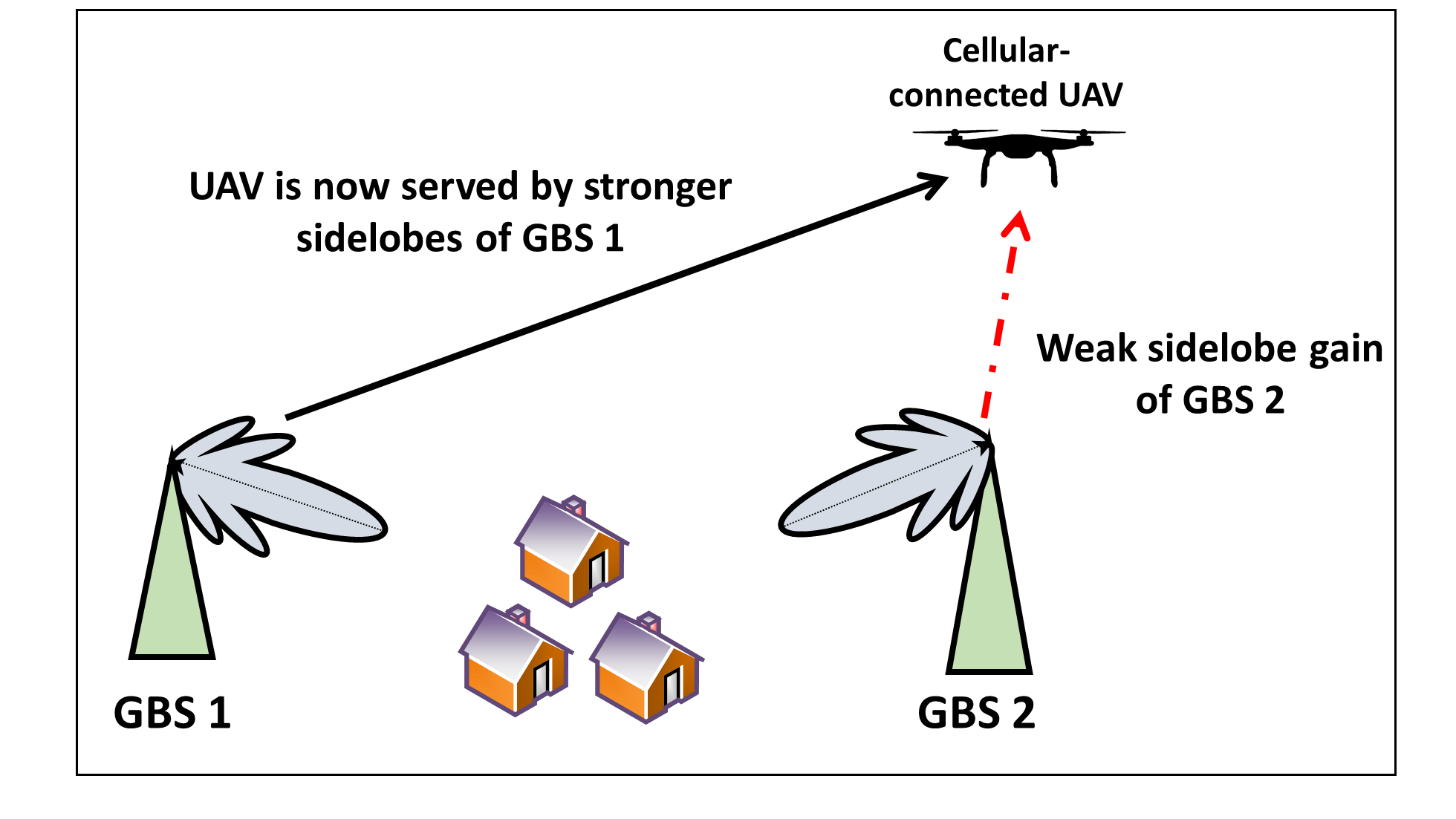}}
		\caption {
		{HO scenario of a cellular-connected UAV moving towards horizontal direction. (a) UAV is associated with GBS 2 due to its higher side-lobe gain than GBS 1. (b) After moving forward, the UAV is now associated with GBS 1 due to its higher sidelobe gain.}}
		\label{fig:HO_explanation}
		\vspace{-.5 cm}
\end{figure}

To fully reap the benefits of UAV deployment, beyond visual line of sight operations are of critical importance where UAVs acting as aerial users, can maintain communication with the ground base stations (GBSs) for command and control (C\&C) purposes in the downlink (DL)\cite{geraci_2018}. UAVs flying in the sky may be served by the sidelobes of base station antennas which provide lower antenna gains\cite{geraci_2018,ramy_antenna}. This poses major challenges on the mobility management (MM) for the cellular-connected UAVs based on reference signal received power (RSRP). The GBS providing the highest RSRP might be located far away from the UAV. An example is shown in Fig.~\ref{fig:HO_explanation}, where a cellular-connected UAV is flying over a rural environment. 
This type of patchy signal coverage of GBSs will result in poor mobility performance such as handover failure (HOF), radio link failure, as well as unnecessary handovers (HOs), called ping-pong events. Apart from these, due to the loss of the C\&C signal, the UAV may collide with a commercial aircraft or even crash into a populated area which might result in hazardous events. Hence, effective MM for providing reliable communications between UAVs and GBSs is of critical importance.

MM techniques for ground user equipment (GUE) in both homogeneous and heterogeneous cellular networks have been studied extensively in the literature~\cite{Simsek2015, karthik2017,lin_mobility}. However, the research in MM for cellular-connected UAVs is still in its infancy. Motivated by this, considering an interference-limited DL cellular network, the main contribution of this paper is a novel a reinforcement learning (RL) based MM technique that can ensure better connectivity at the UAV while also maintaining a reasonable throughput performance for existing GUE. The proposed approach exploits Q-learning \cite{Sutton:1998} for tuning the downtilt (DT) angles of GBSs to meet these goals.In this regard, we assume that the UAV's trajectory is known beforehand and the network can control the RSRP values at different discrete waypoints of the predefined UAV trajectory by changing the DT angles. We also consider the presence of correlated shadow fading~\cite{gudmundson_crr_shad} for the predetermined UAV trajectory. We introduce a reward function that takes into account both RSRP at the UAV and the capacity of the GUE, and we study the performance of different weight combinations associated with both of these performance criteria. Our simulations demonstrate a tradeoff between the RSRP quality at the UAV and the GUE capacity. Moreover, in a scenario, where the network considers maximizing the RSRP values, the number of HOs can be reduced by $40\%$, compared to a scheme where the DT angle is remained fixed.

The rest of this paper is outlined as follows. In Section~\ref{sec:lit_reviw}, we provide a literature review pertinent to MM of cellular-connected UAVs. Section~\ref{sec:sys} describes the system model. We discuss the proposed MM technique is Section \ref{sec:RL_uav}. Related simulations results are presented in Section \ref{sec:simu_rl}. Finally, Section~\ref{sec:Conc} concludes this paper.



\section{Related Works}
\label{sec:lit_reviw}

Integrating UAVs as aerial user equipment with existing GUE has attracted attention in recent years. For instance, authors in~\cite{ramy_saad} presented an analytical framework for co-existing UAV and GUE considering beamforming technique. In \cite{ramy_coverage}, authors provided the upper and lower bounds on the coverage probability of UAVs considering a coordinated multi-point technique. Authors in \cite{challita_2019} investigated the problem of interference-aware optimal path planning of cellular-connected UAVs using deep RL in the uplink (UL) scenario. In~\cite{Ramy_walid_ICC2020}, authors explored the effects of practical antenna configurations on the mobility of cellular-connected UAVs and showed that increasing the number of antenna elements can increase the number of HOs for vertically-mobile UAVs.\looseness=-1   

In \cite{lin_sky} and \cite{lin_field}, real-world industrial experiments were conducted to test the feasibility of integrating UAVs in existing cellular networks. The Third Generation Partnership Project (3GPP) also studied the potential of integrating UAVs as aerial UEs in existing cellular networks and pointed out the challenges in providing reliable mobility support~\cite{3gpp}. According to \cite{denmark_uav_test}, downlink (DL) performance and UL interference can be mitigated by introducing antenna selection or beamforming capabilities at the UAVs. By conducting extensive 3GPP compliant simulations, in \cite{xingqin_ho_2018}, the authors showed that existing cellular networks will be able to support a small number of aerial user equipment (UE) with good mobility support.

In recent work, the authors explored the RL algorithm to maximize the received signal quality at a cellular-connected UAV while minimizing the number of HO~\cite{xingqin_rl_2019}. A novel HO performance optimization algorithm by tuning the values of the 3GPP specified HO parameters in an automated manner was introduced in  \cite{HO_oaram_opt}. An RL-based HO optimization scheme for ground UEs in a 5G cellular network was proposed in \cite{5g_rl}.  

However, none of these prior works~\cite{ramy_saad,ramy_coverage,challita_2019,Ramy_walid_ICC2020,lin_sky,lin_field,3gpp,denmark_uav_test,xingqin_rl_2019,5g_rl} studied the problem of concurrently maintaining good RSRP quality for the UAV and high throughput for GUE while considering practical antenna pattern of the GBS. It is worth noting that the network can control the received power quality of the users by changing the DT angle\cite{dt_effect}. DT angles can be easily and remotely carried out by adjusting the relative phases of antenna elements of an antenna array in such a way that the radiation pattern can be downtilted uniformly in all horizontal directions \cite{edt}. To the best of our knowledge, no prior work has considered tuning DT angles dynamically for effective MM of cellular-connected UAVs. 
\begin{figure}[t]
\centering{\includegraphics[width=0.6\linewidth]{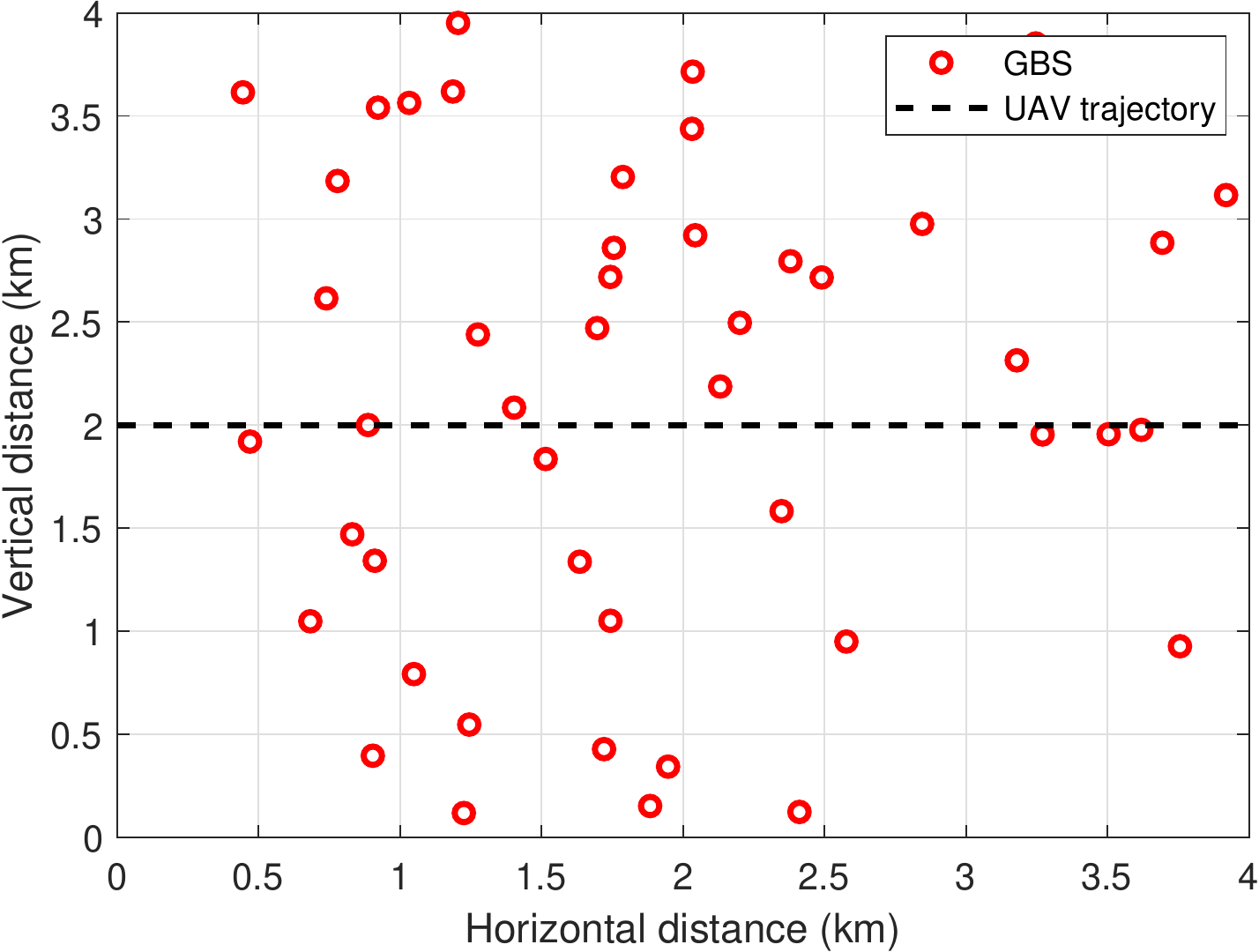}}
    \caption{Illustration of cellular network with linear mobility model.}
    \label{fig:uav_linear_trajec}
 \vspace{-4mm}
\end{figure}

\section{System Model}
\label{sec:sys}
\subsection{Network Model}
Consider the DL of an interference-limited cellular network in which a single UAV acting as an aerial user, is flying along a two dimensional (2D) linear trajectory at a fixed height $h_{\textrm{uav}}$. 
The network consists of $M$ GBSs and $K$ static GUE with similar height $h_{\textrm{GUE}}$. We also assume that all GBSs have equal altitudes $h_{\textrm{GBS}}$ and transmission power $P_{\textrm{GBS}}$. The set of the GUE can be denoted as $\mathcal{K}$ with horizontal coordinates  $\boldsymbol{r}_k=[x_k,y_k,z_k]^T \in \mathbb{R} \textsuperscript{3x1}, {k} \in \mathcal{K}$. We define $\mathcal{M}$ as the set of the GBSs. The GBSs have three sectors separated by 120\degree,~while each sector is equipped with $8 \times 1$ cross-polarized antennas downtilted by the same DT angle $\beta$. We assume that the UAV and the GUE are each equipped with an omnidirectional antenna. We consider the 3GPP antenna radiation model to characterize the antenna radiation at the GBS~\cite{rebato_antenna}.


  
We consider the sub-6 GHz band for the cellular network and, hence, the presence of thermal noise power at the receivers is negligible compared to the interference power. We also assume that the GBSs share a common transmission bandwidth and full buffer traffic is used in every cell. All DL transmissions are scheduled using a round-robin scheduling algorithm, and all the receivers are capable of mitigating the Doppler effect~\cite{rui1,doppler}.
\subsection{UAV Mobility Model}
A simple UAV mobility scenario is considered whereby the UAV travels along a linear trajectory (for instance, through the horizontal x-axis) with constant speed $v$ km/h and a fixed altitude of $h_{\textrm{uav}}$. We consider the linear mobility model due to its simplicity and suitability for UAVs flying in the sky with virtually no obstacle.   
 In Fig.~\ref{fig:uav_linear_trajec}, we provide an illustrative example of the linear UAV mobility model for $v= 120$ km/h in an area of $4\times4$ $\text{km}^2$ with $M=48$.


\subsection{Path-loss Models and GUE Capacity}
The path-loss between a GBS and a UAV plays a significant role in the HO performance due to the RSRP-based cell association. 
For modeling the path-loss, we consider the RMa-AV-LoS channel model specified in 3GPP\cite{3gpp}. The instantaneous path-loss (in dB) under a line-of-sight (LOS) scenario between GBS $m$ and the UAV can be expressed as:
\begin{equation}
\begin{split}
\xi_{{m,u}}^{\text{LOS}}(t)&=\text{max}\big(23.9-1.8\log_{10}(h_{\textrm{uav}}),20 \big)\log_{10}(d_{{m,u,t}})\\&+20\log_{10} \bigg(\frac{40\pi f_c}{3} \bigg)+\chi_{\text{LOS}},
\end{split}
\end{equation}
where  $h_{\textrm{uav}}$ is between 10 m to 300 m and $f_c $ is the carrier frequency, while $d_{{m,u,t}}$ represents the 3D distance between the UAV and GBS $m$ at time $t$. $\chi_{\text{LOS}}$ represents the correlated shadow fading (SF) (in dB) associated with LOS scenario. We will discuss about SF in the next subsection. It is worth noting that the probability of LOS is equal to one if the UAV height falls between 40 m and 300 m \cite{3gpp}. Finally, we can calculate the RSRP (in dB) from GBS $m$ at time $t$ as $\text{RSRP}_{m,u}(t)=P_{\text{GBS}}+ G_{m,u}(t) -\xi_{{m,u}}^{\text{LOS}}(t)$, where $G_{m,u}(t)$ is the antenna gain at the UAV from GBS $m$ at time $t$, which is dependent on the UAV's location and $\beta$~\cite{geraci_2018}.

The path-loss (in dB) observed at GUE $k \in \mathcal{K}$ from GBS $m$ is given by: $\xi_{{k,m}}=A+B\log_{10}({d_{{k,m}}})+C$. Here, ${d_{{k,m}}}$ is the Euclidean distance from GBS $m$ to GUE $k$. $A$, $B$, and $C$ are factors dependent on the carrier frequency $f_c$ and antenna heights \cite{hata}. Then, the received power at GUE $k$ from GBS $m$  can be calculated as ${S_{{k,m}}}=\frac{{P_{\textrm{GBS}}G_{k,m}}}{10^{(\xi_{{k,m}/10)}}}$, where $G_{k,m}$ represents the antenna gain at GUE $k$ from GBS $m$ that depends on the locations of GUE $k$ and GBS $m$~\cite{rebato_antenna}. Note that $G_{k,m}$ will be dependent on $\beta$ as shown in \cite{geraci_2018}. A GUE associates with the GBS that provides the best signal-to-interference ratio (SIR). The SIR between GUE $k$ and GBS $m$ can be calculated as ${\gamma_{k,m}}=\frac{S_{{k},m}}{\sum_{j \in \mathcal{M},j \neq m} S_{{k},j}}$. Assuming that, GUE $k$ is associated with GBS $m$, we can express the achievable data rate per unit bandwidth (bps/Hz) of GUE $k$ using Shannon's capacity as: 
\begin{equation}
C_{k,m}=\frac{\log_{2}(1+\gamma_{k,m})}{N^m_{\mathrm{GUE}}}, 
\label{eqn:rate}
\end{equation}
where ${N^m_{\mathrm{GUE}}}$ is the number of GUEs associated with GBS $m$. Since we are considering round-robin scheduling with full buffer traffic, the available resources will be equally divided among the GUEs associated with a particular GBS. Assuming that the set of GUEs associated with GBS $m$ is $\mathcal{K}_m$, the total GUE capacity can be calculated as $C= \sum_{m \in \mathcal{M}}\sum_{k \in \mathcal{K}_m} {C_{k,m}}$. 

\subsection{Shadow Fading Model}





SF also known as medium-scale fading, plays a critical role in characterizing the received power at the receiver. It is caused mainly due to the presence of large obstacles within a wireless link \cite{Goldsmith:2005:WC:993515}. SF is typically modeled as an independent Gaussian random variable with zero mean and standard deviation $\sigma$. According to \cite{3gpp},  $\sigma$ (in dB) can be expressed as $ \sigma= 4.2\exp({-0.0046~{h}_{\text{uav}}})$.
However, the SF values of consecutive waypoints of a UAV trajectory might have some similarities or correlation due to high probability of LOS in the GBS-to-UAV link. Based on empirical measurements, the work in~\cite{correlated_shadowing} first proposed the common analytical model for auto-correlation among the SF values which assumes that SF is a first-order auto-regressive process where the auto-correlation between the SF values at two points separated by distance $\Delta$ is given by \cite{Goldsmith:2005:WC:993515}:
\begin{equation}
    R(\Delta)=\sigma^2  \rho^{\frac{\Delta}{X_c}},
    \label{eq:auto_corr}
\end{equation}
where, $\Delta$ is the distance between the two points, $ \rho$ is the correlation coefficient, and $X_c$ is the decorrelation distance \cite{Goldsmith:2005:WC:993515}. Here, we set $X_c=100$ m and $\rho=0.82$~\cite{Goldsmith:2005:WC:993515}. We first generate independent SF values with zero mean and $\sigma=1$ for each waypoint and then use Cholesky factorization to generate the correlated SF values from $R(\Delta)$~\cite{Klingenbrunn_corr_shadow}.

\subsection{Handover Procedure}
Throughout the flight duration, the UAV can measure the RSRPs from all the adjacent GBSs at subsequent measurement gaps using (1). We consider a HO mechanism that involves a HO margin (HOM) parameter, and a time-to-trigger (TTT) parameter, which is a time window which starts after the following HO condition (A3 event \cite{3gpp.36.331}) is
fulfilled. 
\begin{equation}
\text{RSRP}_{{j}}>\text{RSRP}_{{i}}+m_{\text{hyst}},
\label{a3_event}
\end{equation}
where $\text{RSRP}_{{j}}$ and $ \text{RSRP}_{{i}}$ are the RSRP measured from the serving GBS $i$ and target GBS $j$, respectively and $m_{\text{hyst}}$ is the HOM set by the network operator. The UAV does not transmit its measurement report to its current serving GBS before the TTT timer expires~\cite{karthik2017}. 


\section{RL-based Mobility Management }
\label{sec:RL_uav}
In this section, we will describe our proposed RL-based MM framework. The aim is to determine the optimal sequence of DT angles for each discrete waypoint of the predefined UAV trajectory for maintaining good RSRP quality at the UAV while maintaining good GUE rate performance. This problem can be solved by non-linear optimization techniques~\cite{Bertsekas/99}. However, optimization-based techniques require the exact knowledge of
the modeling parameters, which might not always be available. Moreover, even with the perfect information of all relevant parameters, such an
optimization problem is NP-hard~\cite{np_hard_tilt} and difficult to solve efficiently due to the intractability in GBS antenna radiation pattern, HO process, and channel models considered in this research.

Thus, we consider RL algorithm for solving this sequential decision problem in which an agent tries to maximize its cumulative rewards by interacting with an unknown environment through time~\cite{ai_survey}. The agent gets an immediate reward or punishment for taking any action and tries to maximize the total expected future reward. Usually, the environment is modeled as a Markov decision process (MDP) which is characterized by a tuple $(\mathcal{S},\mathcal{A},P,R)$, where $\mathcal{S}$ is the set of states,  $\mathcal{A}$ is the set of actions, $P$ is the state transition probability function, with $P(s'|s,a)$ denotes the probability of moving to the next state $s' \in \mathcal{S}$ from the current state $s \in \mathcal{S}$ after taking action $a \in \mathcal{A}$, and $R:\mathcal{S} \times \mathcal{A} \rightarrow \mathbb{R}$ is the immediate reward received by the agent. An MDP can be solved with this information to obtain the optimal policy, i.e., the action to take at each state which will maximize the expected sum of discounted future rewards.




A branch of RL known as model-free RL algorithms can learn optimal policies in finite MDPs without explicit knowledge about the environment modeling, i.e., $P$ and $R$. They are typically simpler and more flexible to implement than their model-based counterpart since the dynamics or the model of the environment is not required to be known in prior. In a typical cellular network, to obtain the perfect knowledge of the environment, a significant amount of information has to be exchanged between the network components and the core network, which is not always possible. Hence, for our proposed MM scheme, we use a well-known model-free RL algorithm known as Q-learning~\cite{Sutton:1998}.
Q-learning approximates a value function of each state-action pair through several iterations and learns the optimal policy by using this function.

In our proposed algorithm, we divide the predefined UAV trajectory into discrete states whereby, at each state $s$, the network can perform an action $a$ and then the UAV moves to next state $s'$. The reward for taking action $a$ in state $s$ is denoted as $R(s,a)$. The learned action-value $Q(s,a)$ for performing action $a$ in state $s$ is updated using the following rule at~each~iteration:
\begin{equation}
{Q(s,a)\leftarrow(1-\alpha)Q(s,a)+\alpha\big[R(s,a)+\lambda\max\limits_{a'\in \mathcal{A}}Q(s',a')\big]},
\label{eq:q_fn}
\end{equation}
where $\lambda \in [0,1)$ is the discount factor, $a'$ is the action that will be taken in the next state $s'$, and $\alpha \in [0,1]$ is the learning rate. After the iterative process, the agent will eventually learn the optimal Q-values for each state-action pair, $Q^*(s, a)$ over time. Then, the optimal policy can be obtained by acting greedily in every state by the following equation.
\begin{equation}
\label{eq:Update_eqn}
\pi^*=\argmax\limits_{a \in \mathcal{A}} Q^*(s,a).
\end{equation}

We discuss the modeling of state, action, and reward associated with our MM scheme in the following subsection.

\subsection{Simulation Setup and Training}
\subsubsection{State Representation}
We take the predetermined linear trajectory of the UAV of duration $L$ and divide into $\delta$ equal segments where $\delta=\frac{L}{n}$. Here $n$ represents the measurement gap. Note that the given trajectory might already be calculated offline in order to maximize some performance criteria. We denote each segment or waypoint $(x_{\text{uav}}, y_{\text{uav}}, h_{\text{uav}}) $ as a unique state. Hence, the number of available states in our setup is $\delta$. 
\subsubsection{Action Representation}
The network acting as the agent can choose $\beta$ in the range $[-2,12]$ with $2^\circ$ resolution. In other words, the number of available actions is eight. The network can change the values of $\beta$ by changing the phase of each antenna element remotely during each $n$.

\subsubsection{Reward Model}
To obtain the desired performance, we define a reward function that focuses on maximizing the GUE sum-rate and the RSRPs at the UAV simultaneously along the route. During a flight, the agent needs to tune $\beta$ of the GBSs while maintaining both of these goals. The weighted combination of RSRP and GUE capacity for taking action $a$ at state $s$ can be expressed as:
\begin{equation}
R(s,a)=w_{\text{rate}}\times C_{s'}+w_{\text{RSRP}}\times\text{RSRP}_{s'},
\label{eq:reward}
\end{equation}
where $C_{s'}$ and $\text{RSRP}_{s'}$ are the rate of the GUE and the RSRP from the serving GBS at following state $s'$, respectively. $w_{\text{rate}}$ is the weight associated with the GUE rate and $w_{\text{RSRP}}$ represents the weight of the RSRP. The weights are chosen such that $w_{\text{rate}}+w_{\text{RSRP}}=1$.

\begin{algorithm}[!h] \small
\caption{Q-learning for UAV mobility management.}
\label{alg:Alg2}
\begin{algorithmic}[1]
\STATE  \textbf{Input}: UAV trajectory, cellular network, $\epsilon=1$, $\alpha$, $\gamma$, $Q=\mathbf{0}_{\delta \times 8}$
\STATE \textbf{repeat} (for each iteration):
\STATE \hspace{0.3cm} \textbf{if} $\epsilon$ $\geq$ $\epsilon_{\rm min}$
		\STATE \hspace{0.6cm} $\epsilon$=$\epsilon$ $\times$ $\eta$ \textbf{end if}\\
\STATE \hspace{0.3cm}\textbf{repeat} (for each segment of the UAV trajectory):
\STATE \hspace{0.6cm} choose action $a$ using $\epsilon$-greedy policy:
\STATE \hspace{0.6cm} take action $a$, calculate $R(s,a)$ using \eqref{eq:reward}
\STATE \hspace{0.6cm} update $Q$-value using \eqref{eq:q_fn} and by calculating the action\\ \hspace{0.6cm} which maximizes $Q$-value in the next state
\STATE \hspace{0.6cm} $s \leftarrow s'$
\STATE \hspace{0.3cm} \textbf{until} $s$ is terminal
\STATE  \textbf{Output}: $Q^*$ table
 \vspace{-1.25mm}
\end{algorithmic}

\end{algorithm}

\subsubsection{Training Process}
Our offline Q-learning training process consists of multiple iterations or epochs, each with a number of steps $\delta$. At each step, the network tries to obtain the highest reward while also checks for other actions that can improve the estimated future reward. To overcome this exploration-exploitation dilemma, we adopt $\epsilon$-greedy method \cite{watkins}. The main idea is to choose a random number from $[0,1]$ and check whether it is smaller than $\epsilon$. If so, the agent takes a random action; otherwise, it goes with the action that has the highest $Q$-value. We start our first iteration with $\epsilon=1$ and then reduce $\epsilon$ by multiplying a decay factor $\eta=0.99$ at each iteration. Each iteration starts with $s=0$ (first waypoint of the UAV trajectory) and finishes when UAV reaches the final location. The pseudo-code of the proposed Q-learning process is presented in Algorithm~\ref{alg:Alg2}.

It is worth noting that the actions chosen by the agent will impact the RSRP values, which will eventually trigger the HO condition at each state. We count the number of HOs for a given trajectory as follows. At the beginning of the trajectory ($s=0$), the RSRPs from all the available GBSs is calculated and the UAV associates with the one providing the highest RSRP. We denote the current cell as $\mathcal{C}_{\text{curr}}$. For the next waypoint or $s=1$, $\mathcal{C}_{\text{curr}}$ remains the same. At $s=2$, the RSRPs from all GBSs is calculated again and the GBS with the highest RSRP at $s=2$ is denoted as $\mathcal{C}_{\text{next}}$.

If $\mathcal{C}_{\text{next}}$ is different that $\mathcal{C}_{\text{curr}}$ and the measured RSRP of $\mathcal{C}_{\text{next}}$ is higher than that of $\mathcal{C}_{\text{curr}}$ by HOM, we calculate the RSRP of $\mathcal{C}_{\text{next}}$ and $\mathcal{C}_{\text{curr}}$ after travelling TTT s between $s=1$ and $s=2$. If the RSRP of $\mathcal{C}_{\text{next}}$ is higher than that of $\mathcal{C}_{\text{curr}}$ by HOM, at the moment when TTT ends, HO to $\mathcal{C}_{\text{next}}$ takes place. Otherwise, the UAV keeps associated with $\mathcal{C}_{\text{curr}}$. The same procedure is repeated at each waypoint until UAV reaches the last state.



\subsubsection{Analysis of the Algorithm}
In our proposed algorithm, the number of states or the discrete waypoints is finite. The reward function can take values in the range $[0,1]$ and the set of allowed actions is also finite. Moreover, a decision epoch or iteration finishes when the UAV reaches the final location. Hence, our MM method is an MDP. If we consider $\lambda < 1$ and $\alpha \in [0,1]$, our proposed Q-learning based MM technique will converge to an optimal action-value function given that $Q$-values get sufficient number of updates. The complexity of the Q-learning algorithm is $\mathcal{O}(|\mathcal{S}||\mathcal{A}|)$, where $|\cdot|$ denotes the cardinality of a set.


\section{Simulation results}
\label{sec:simu_rl}

\begin{figure}[t]
\centering{\includegraphics[width=0.75\linewidth]{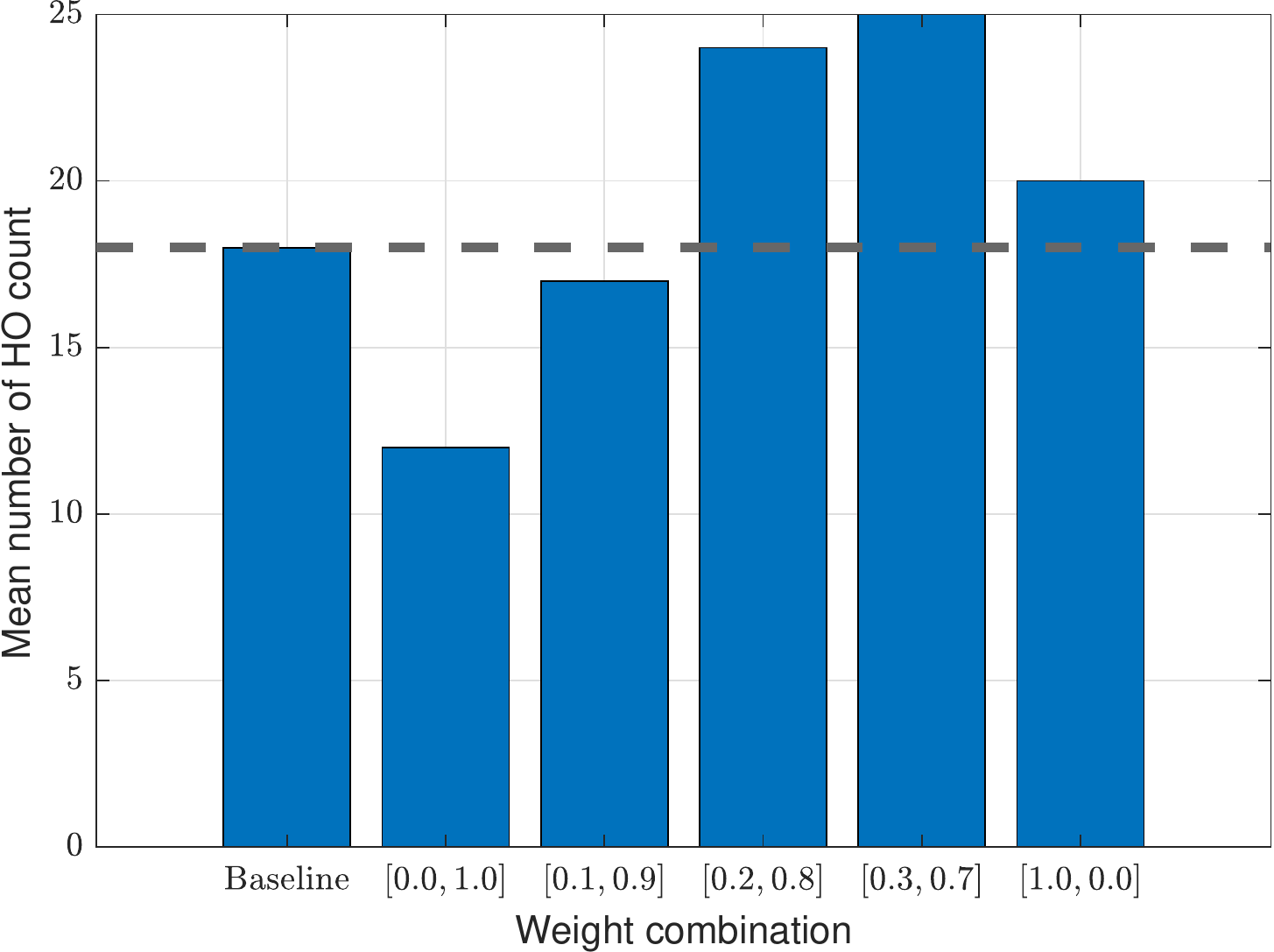}}
    \caption{Average number of HOs per flight for various weight combinations.}
    \label{fig:ho_count_bin}
    \vspace{-2mm}
\end{figure}

\begin{table}[t]
	\vspace{-2mm}
\centering
\renewcommand{\arraystretch}{1}
\caption {Simulation parameters.}
\scalebox{1}
{\begin{tabular}{lc}
\hline
Parameter & Value \\
\hline
$P\textsubscript{gbs}$ & 46 dBm  \\ 
$h\textsubscript{uav}$ & 100 m\\ 
$h\textsubscript{gbs}$ & 35 m\\ 
$h\textsubscript{gue}$ & 1.5 m\\ 
${f_c}$ & 1.5 GHz\\ 
$M$ & 64\\ 
$K$ & 320\\ 
$v$ & 120 km/h\\
$n$ & 200 ms\cite{karthik2017}\\
TTT, HOM & 160 ms, 3~dB\\
\hline
\end{tabular}}
\vspace{-3mm}
\label{tab}
\end{table}
In this section, we evaluate the performance of our proposed RL-based MM technique. We consider terrestrial networks containing 64 GBSs and 320 GUE in a square area of $4 \times 4$ $\text{km}^2$, where the GBSs and the GUE are placed randomly. We consider that the UAV is flying at a speed of 120 km/h in a straight line and the route faces correlated SF. The flying time of the UAV is considered to be 120 seconds. Here, we ignore the take-off and landing time of the UAV as well as the time to reach the altitude $h_{\textrm{uav}}$. Note that our method can allow more flying time at the expense of higher computational complexity. For Q-learning training purpose, we consider $\alpha=0.8,$ and $\lambda=0.9$. The Q-learning algorithm is trained offline for 1500 iterations or epochs. We consider 100 random networks for simulation purposes and then compute the average to study the performance of our proposed method. Note that at each realization, the UAV trajectory remains the same; the underlying positions of the GUE and GBSs as well as SF values associated with trajectory change. For each random realization, the RSRP samples associated with each discrete waypoint are linearly transformed to the interval [0,1]. We also normalize the GUE rates associated with each action or $\beta$ for each random network. The simulation parameters and their default values are listed in Table \ref{tab}.

We consider different weight combinations of RSRPs and GUE rates to test the performance of our proposed MM technique and denote weight vector $\mathbf{w}=[w_{\text{rate}},~w_{\text{RSRP}}]$ 
to represent different weight combinations. For instance, $[1.0,0.0]$ means that the values of $w_{\text{rate}}$ and $w_{\text{RSRP}}$ are $1.0$ and $0.0$, respectively. To compare the performance of our approach, we also consider a baseline in which the value of $\beta$ is kept fixed at $6^\circ$ as done for the RMA-AV scenario in~\cite{3gpp}.
Since

Fig.~\ref{fig:ho_count_bin} shows the mean number of per-flight HOs completed by the UAV for different weight combinations. We can observe that $[0.1,0.9]$ and $[0.0,1.0]$ provide $15\%$ and $40\%$ fewer average HOs, respectively than the baseline scheme where $\beta$ is kept fixed. On the other hand, a weight vector $[1.0,0.0]$ scheme yields $11.1\%$ more average HOs compared to the baseline. This is because the more we emphasize the RSRP values, lower values of $\beta$ are selected which will steer the main beams of the GBSs towards the UAV. This leads to smoother signal coverage in the sky which results in lower HO count. $[{0.2},{0.8}]$ and $[0.3,0.7]$ schemes vary $\beta$ sharply during the flights which create patchy GBS coverage from the UAV's perspective and hence, causes frequent HOs.

\begin{table}[h]
\centering
\renewcommand{\arraystretch}{1}
\caption {DT angle for different weights.}
\label{tab_angle}
\scalebox{1}
{
\begin{tabular}{lc}
\hline
Weight vector & mean $\beta$ \\
\hline
$[{0.0},~{1.0}]$ & $-2^{\circ}$ \\
$[{0.1},~{0.9}]$ & $6.04^{\circ}$\\
$[{0.2},~{0.8}]$ & $11.97^{\circ}$ \\
$[{0.3},~{0.7}]$ & $11.99^{\circ}$ \\
$[{1.0},~{0.0}]$ & $12^{\circ}$ \\
\hline
\end{tabular}}
\end{table} 

In Table~\ref{tab_angle}, we present the per flight mean $\beta$ selected by the network for different weight combinations. We observe that when more emphasis is placed on the GUE rate, the GBSs will then choose higher values for $\beta$. This is consistent with the fact that higher DT angles will decrease the inter-cell interference among the GBSs and will steer the main beam of the antenna towards the GUE with higher gains.

\begin{figure}[t]
\centering{\includegraphics[width=0.75\linewidth]{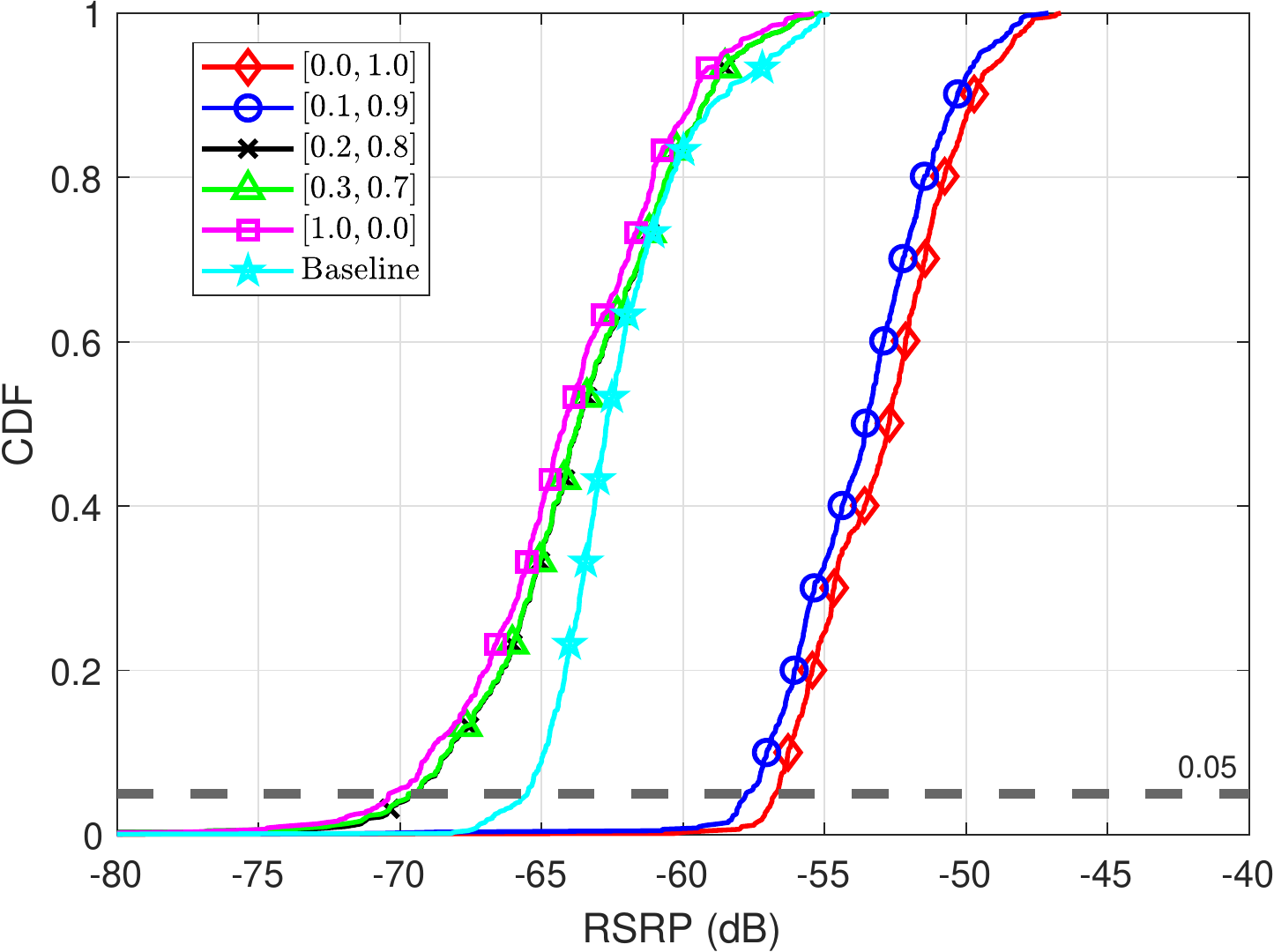}}
    \caption{CDF of RSRP experienced by the UAV for various weight combinations.}
    \label{fig:rsrp}
    \vspace{-2mm}
\end{figure}

In Fig.~\ref{fig:rsrp}, we show the cumulative distribution function (CDF) of RSRPs experienced by the UAV in a flight. The weight vector combination $[{1.0},{0.0}]$ focuses only on the GUE rate and hence, ignores the RSRP performance at the UAV completely. As such, this provides the lowest RSRP for the UAV. On the other hand, $[{0.0},{1.0}]$ provides the highest RSRP. The difference between the fifth percentile RSRP values associated with these two extremes is around 14 dB. $[{0.0},{1.0}]$ also provides about 10 dB higher fifth percentile RSRP than the baseline scheme, whereas $[{0.1},{0.9}]$ provides about 9 dB higher fifth percentile RSRP. The RSRP values associated with the other two weight vectors are very close to each other since they maintain the same average $\beta$ during the flight duration.

\begin{figure}[t]
\vspace{-1mm}
\centering{\includegraphics[width=0.75\linewidth]{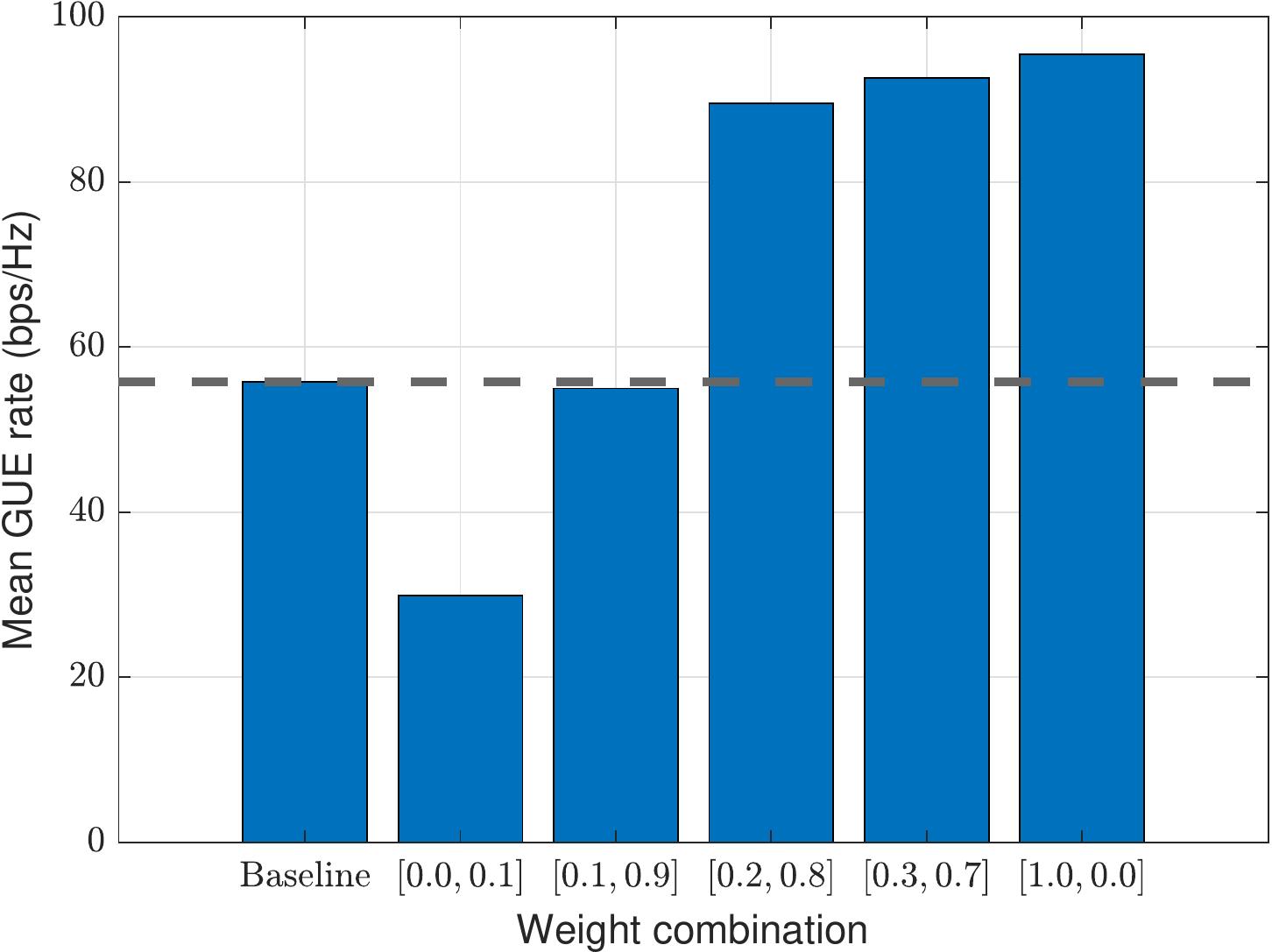}}
    \caption{Average total sum-rate (bps/Hz) of the GUE per flight for various weight combinations.}
    \label{fig:gue_rate_bin}
    \vspace{-2mm}
\end{figure}

In Fig.~\ref{fig:gue_rate_bin}, we show the per flight mean GUE rate associated with different weight vectors. As expected, $[{1.0},{0.0}]$ provides the best GUE rate while its counterpart $[{0.0},{1.0}]$ provides the lowest one. We can see that $[{0.1},{0.9}]$ performs very close to the baseline. 
Overall this weight combination finds a balance between signal quality at the UAV and providing DL coverage to GUE. Moreover, choosing  $\mathbf{w}=[0.1,0.9]$ provides a smaller HO count compared to the baseline scheme.

\section{Conclusion}
\label{sec:Conc}

In this paper, we have proposed an RL-based MM framework in a DL cellular network for ensuring better connectivity for both the cellular-connected UAV and the GUE. By exploiting Q-learning, we have provided a flexible technique for finding a balance between these two contradictory goals. The network can trade-off the received signal strength at the UAV with the GUE rate by adjusting the respective weights of these quantities in the reward function and thus by tuning the DT angles accordingly. Our simulation results have demonstrated that the proposed approach can reduce the number of HOs while maintaining good connectivity to the UAV and the GUE, compared to the scenario where the DT angle is always kept fixed. 
Future extensions can include the case in which multiple UAVs are considered as well as scenarios where the GUE are mobile. In addition, the presence of intelligent reflective surfaces that are facing upward is an interesting future work. 


\bibliographystyle{IEEEtran} 
\bibliography{ref.bib}
\end{document}